# Signatures of Surprising Diffusion-Entropy Scaling across Pressure Induced Glass Transition in Water

Saumyak Mukherjee and Biman Bagchi*

*Solid State and Structural Chemistry Unit, Indian Institute of Science, Bangalore, India*

*Corresponding author electronic mail: bbagchi@iisc.ac.in

## Abstract

Because of the negative inclination of the solid-liquid phase separation line in water, ice Ih melts on compression. On further increase in pressure the liquid water transforms into a high density metastable glassy state, characterized by a rapid approach to zero diffusion coefficient and an absence of any crystalline order in the static structure factor. The vitrification is found to occur even at high temperatures (T > 250 K). We study this glass transition process at four temperatures (80 K, 250 K, 300 K and 320 K). The transition pressure increases with increase in temperature, as expected. Interestingly, we find that the total entropy of the system exhibits a sharp crossover near the glass transition pressure *where the diffusion of water goes to zero*. The diffusion coefficient shows an exponential dependence on the properly defined excess entropy. In an interesting result not reported before, *we find a pressure induced realignment of water molecules resulting in two well separated peaks in the O-O-O angle distribution among neighbouring molecules. The difference between the positions of these two peaks undergoes a sharp change at the vitrification pressure suggesting that it can serve as an appropriate order parameter to detect the glass transition point*.

**Graphical Abstract**

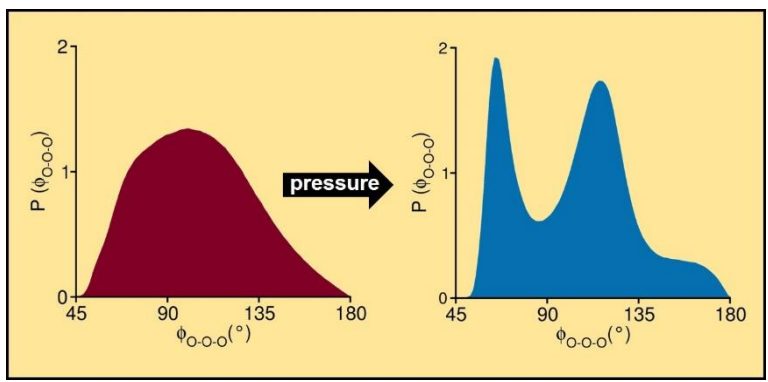

# I. Introduction

Glassy phases of water and their preparation techniques have fascinated researchers, both experimentalists and theoreticians, over many decades.[1-10] Amorphous ice phase is predicted to be the major state of water found in extra-terrestrial space. The low temperature physics of water discusses a number of amorphous states, besides a complex array of crystalline phases.[8, 11-17] Water has a rich and complicated phase diagram with 19 experimentally observed crystalline polymorphs (and several others observed in computer simulations).[18-22] Cooling liquid water beyond its freezing temperature (273 K) is possible till ~230 K, after which spontaneous crystallization sets in. Several simulation studies propose that in this supercooled state, liquid water may have two distinct phases, namely high (HDL) and low (LDL) density liquid, with a critical point.[5, 9, 14, 15, 23-25] A recent work by Nilsson and coworkers has reported experimental observation of HDL-LDL transition.[26] Further below in the temperature scale (beyond 130 K) glassy states including low (LDA), high (HDA) and very high (VHDA) density amorphous ice forms are found.[5-7, 10] The methods of preparation of these phases (sudden temperature quenching, surface deposition of vapour etc.), and the mutual transitions between them have fascinated researchers over a long time.[5, 8]

The glassy phases of water have been reproduced in experiments, and also in computer simulation.[4-7, 9, 27, 28] What we lack is a microscopic understanding of the relative orientation of the water molecules in the glassy phases. Even at high density created by applying high pressure, the long range electrostatic interactions between two water molecules continues to make a major contribution, larger than the contribution from the Lennard-Jones type short range interaction. At extreme high pressure of course, we expect the short range repulsive contribution to dominate. However, over a vast range of pressure, from 10 kbar to 150 kbar or even more, the interplay between these two types of interaction play a major role in dictating the relative stability of the phases and their microscopic structures.

It is fair to say that despite a large body of work, there is a lack of an in-depth understanding of the free energy landscape of ice that produces so many polymorphs, each characterized by at least one local minimum in a properly defined order parameter plane. In fact, one needs to realize that the formation of glassy phases could be explained in terms of the Ostwald Step Rule (OSR) that shows how a glass can be formed because of the proximity of its free energy minimum to the starting phase.[29, 30] However, any analysis employing OSR needs a proper order parameter. It is observed that many of these transitions are accompanied by large changes in density and entropy. What is urgently needed is a microscopic order parameter. Such an order parameter is hard to resolve in a liquid to glass transition. We are not aware of any study that proposes such an order parameter that describes the ice-to-liquid and ice-to-glass transitions.

Freezing-melting transitions of ice and water have drawn renewed interest both, in bulk and confined environments.[31-36] In computer simulations, it is hard to freeze water into crystalline ice at ambient conditions, although accomplished recently.[31, 37, 38] The melting of ice is relatively easy, by increasing temperature under ambient pressure.[32, 39, 40] Studies by Ohmine and coworkers demonstrated that defects (3- and 5-coordinated hydrogen bonded water molecules, and 5 or 7 membered rings) embedded in the 4-coordinated hydrogen-bonded network of water molecules form an entangled state, which play a crucial role in promoting the melting transition.[32] The local structures thus formed are important for the stability of ice polymorphs.[38] In a study by Saito $et\ al.$ it was shown that fragmentation of high-density liquid clusters containing a large number of 3- and 5-coordinated defects lead to anomalous properties of low-temperature water.[41] Subsequently, Saito and Bagchi studied the role of these defects in the glass transition of low-temperature water.[42] Since ice Ih is of lower density than water, pressure-induced melting poses interesting possibilities that we explore here.

The effect of pressure on ice was initially studied in the pioneering experimental work by Mishima $et\ al.$[8] They observed the formation of amorphous ice from ice Ih at T = 77 K and P = 10 kbar. Subsequently, in simulation studies, the amorphization of ice Ih at T = 80 K and P = 13 kbar was reported by Klein and coworkers using the TIP4P water model.[1-3]. Besides this transition, an interesting development was observed at very high pressures (150 kbar), where signatures of crystalline order started to emerge from the disordered phase. Several recent simulation studies have also investigated these high pressure phase transitions of water using multiple water models.[22, 27, 43, 44] These studies reveal the existence of several exotic phases of water, otherwise hidden in the metastable regions of the phase diagram and not readily observed in experiments. Such transitions ultimately lead to ice VII at very high pressures. A recent experimental study indeed suggested that this scenario could be correct, and what so long has been considered as a very high density amorphous (VHDA) ice phase actually consists of a sequence of ordered phases leading all the way to ice VIII as pressure is increased at 100 K.[17, 45] Experiments have probed the fate of ice at pressures, as high as 2100 kbar.[13] However, exact transition point of this pressure induced phase transition is difficult to detect.

The relation between pressure and temperature on phase coexistence is given by the slope of the P-T coexistence line in the phase diagram of water.[18] It is described quantitatively by the well-known Calusius-Clapeyron equation.[46] An important parameter that determines the gradient of the coexistence line is the difference between volumes of the two phase. Therefore, the nature of the conversion process between the various forms of crystalline and liquid water is intrinsically dependent on their densities. While often not stated clearly, the latent heat $L$ (related to the entropy change) and volume change are not independent because the chemical potential must be the same in the two phases. This relationship is brought out most clearly by the density functional based theory of freezing of Ramakrishnan and Yussouff.[47]. This theory (R-Y theory) gives the dependence of fractional density change on freezing and the entropy change on the two and three particle direct correlation functions and the symmetry of the lattice encoded in terms of reciprocal lattice vectors. However, freezing of water into ice Ih is quite difficult to explain within the R-Y theory because the density change is negative which is due to the open framework structure of ice Ih. We do not have accurate values of two and three particle direct correlation functions for water needed to describe freezing of liquid water.

The comparison provides insight because the simpler (than water) radial intermolecular potential dominated by the repulsive part at short distances dictate the freezing transition. The question then naturally arises: what happens to the solid phases of ice at very high pressure where the open framework of ice Ih is bound to become unstable and water molecules are forced close to each other, to experience the Lennard-Jones like potential that forms the core of every intermolecular potential of ice? Does this shorter (than polar) range part of water-water potential dictate freezing/melting transitions at high pressure and high density?

In this work, we explore pressure induced phase transition of water from a microscopic perspective, and establish an order parameter description of such transitions. Our main focus lies on the structural and thermodynamic changes in a system as a function of pressure. We scan a wide range of pressures, starting from 10 kbar to as high as 150 kbar. Following the sharp first order phase transition from crystalline ice Ih to high density supercooled liquid (or very high density amorphous ice at low temperature), we observe a subsequent less sharp transition to crystalline order at higher pressure. We characterize this transitions by structural, dynamic and thermodynamic parameters. We find that the difference between the two peaks in O-O-O angle distributions among neighbouring water molecules serves as an excellent order parameter to study the recrystallization transition. This agrees well with the observed transition sin diffusion coefficient and total entropy of the system. O-O radial distribution function (RDF) suggests the possible presence of several entangled crystalline polymorphs in the studied pressure regime. Both O-O RDF and O-O-O angle distributions capture the incipient transformation of the system to crystalline ice VII. Although, crystallization is not observed in the time scale of our simulations. Complexity of these polymorphic structures, and their mutual transformations, suggest a possible connection with two fundamental theoretical formulations: Ostwald Step Rule and Random Spin Ising Model, the correlation between which have been discussed in a recent work.[29]

## II. Glass transition at 250 K, 300 K, and 320 K

We start our simulations on an ice Ih crystal consisting of 1024 TIP4P/Ice water molecules at 250 K, 300 K, and 320 K and at pressures ranging from 0 to 150 kbar. Upon compression, the crystal melts to liquid water because of the negative slope of the solid-liquid coexistence line in the phase diagram of water. On further compression (p > 30 kbar), the liquid water experiences a phase transition. The ensuing phase of water is characterized by zero diffusivity, and no apparent crystalline order in the static structure factor (**Figure S1**). Hence, the molten liquid undergoes a vitrification at pressures greater than 30 kbar.

### A. Density, Diffusion and Entropy Changes

In **Figure 1**, we plot the pressure dependence of density ($\rho$) of the system at these 3 temperatures. With increased compression, the density of the system increases nonlinearly. At 150 kbar, the value of $\rho$ is 1.79 g cm$^{-3}$, which is similar to the density of ice VII at that condition (observed from our simulation). The packing fraction of the system at this pressure is significantly high. With increase in pressure, the compressibility of the system decreases substantially, which results in the nonlinear increase of density. At very high pressures, the system reaches a state where the density becomes indifferent of temperature. Here, the system is no longer in a molten liquid state, but behaves like a rigid solid.

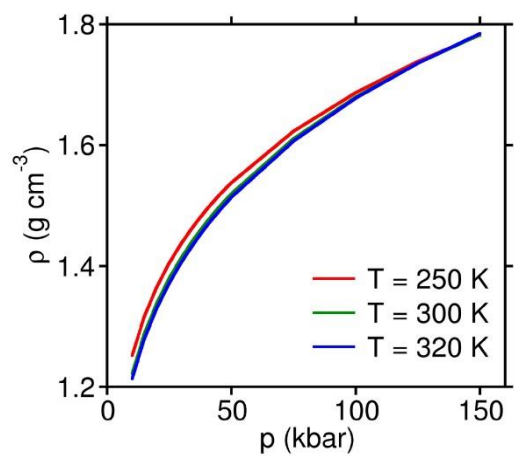

**Figure 1.** Change in the mass density (ρ) of system with pressure (p) at 250 K (red), 300 K (green) and 320 K (blue). The changes are almost similar at all the 3 temperatures. At 150 K, the value of ρ is similar to that of ice VII at that condition. The compressibility of the system is significantly reduced owing to the extreme compression, which is reflected in the nonlinearity in the p-dependence of ρ.

Owing to the gradual increase of density, the exact point of phase transition cannot be captured from **Figure 1**. However dynamics of the system, such as self-diffusion coefficient (D) of the water exhibits a more distinct change at the transition pressures. D is calculated from the time derivative of the mean square displacement (MSD) of water [**Eq.(1)**].

$$D = \frac{1}{6}\frac{d}{dt}\left\langle \Delta r^2(t) \right\rangle \quad (1)$$

At the said temperatures (250 K, 300 K and 320 K) the molten state is a compressed liquid. This is clearly reflected in the non-zero values of diffusion coefficient. As observed in **Figure 2a**, after a certain pressure D becomes zero. The structure factor of the phase of water beyond this pressure does not exhibit any crystalline order (**Figure S1**). Hence, based on the dynamics and the structure of the system, it can be recognized as a glassy phase. At the temperatures studied, the transition pressures ($p_{trans}$) at which D becomes zero are ~30 kbar, ~50 kbar and ~65 kbar respectively (**Figure 2a** inset).

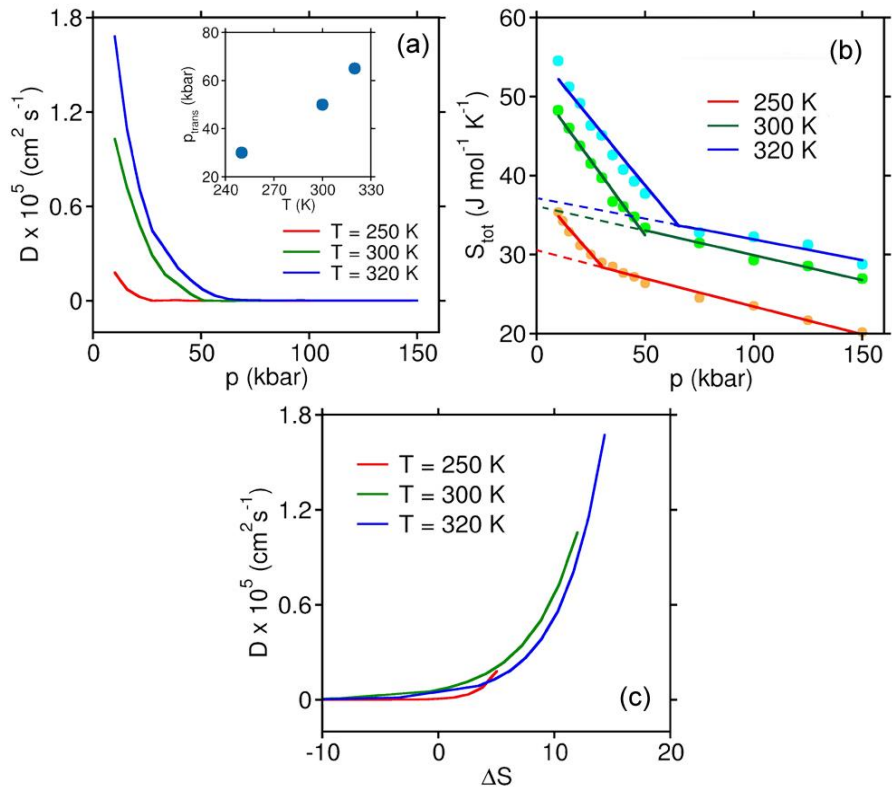

**Figure 2.** (a) Self-diffusion coefficient (D) of water molecules in the system is plotted as a function of pressure at the 3 temperatures (250 K (red), 300 K (green) and 320 K (blue)). Compression results in a liquid to solid transition, whereby the value of D becomes zero beyond the phase transition pressure ($p_{trans}$). The value of $p_{trans}$ at the 3 said temperatures are 30 kbar, 50 kbar and 65 kbar respectively (plotted in the inset). (b) Total entropy ($S_{tot}$) of the system is plotted as functions of pressure. At all the three temperatures, $S_{tot}$ shows a change in slope at the transition point. The two branches are fitted to linear functions, the intersection of which yields $p_{trans}$. (c) Diffusion coefficient (D) is plotted as a function of the excess entropy ($\Delta S$) in the liquid phase, with respect to the glass formed after the transition. D scales exponentially with $\Delta S$. $\Delta S$ is equivalent to the configurational entropy of the liquid state.

It is often observed that diffusion coefficient (D) is strongly correlated to the entropy (S) of a system.[48-50] These are two seemingly unrelated quantities. While diffusion is a transport property, entropy explores the thermodynamics of the system. However, one can express diffusion as a function of entropy by the well-known expressions of Rosenfeld relation [51, 52] and Adam-Gibbs relation.[53] For the solid phases (crystalline and amorphous), a total quantum treatment in terms of the harmonic oscillator partition function ($Q$) yields accurate entropy of the system.[54]

$$S = k_B \ln Q + \beta^{-1} \left( \frac{\partial \ln Q}{\partial T} \right)_{N,V}$$
$$= k_B \int_0^\infty d\nu g(\nu) \left[ \frac{\beta h \nu}{e^{\beta h \nu} - 1} - \ln\left(1 - e^{-\beta h \nu}\right) \right] \quad (2)$$

Here, $k_B$ is the Boltzmann constant and $\beta = (k_B T)^{-1}$. $g(\nu)$ is the spectral density (density of states) at frequency $\nu$ and $h$ is the Planck's constant. However, in the presence of a diffusive component in the liquid state, a harmonic oscillator treatment is not enough. Hence, we use the 2PT method introduced by Goddard and coworkers to calculate the entropy of the system.[55-58] The resultant graphs are plotted in **Figure 2b**. The total entropy of the system ($S_{tot}$) decreases with the increase in pressure. At all the studied temperatures, the change in total entropy exhibits a distinct crossover which results in two branches with different gradients. We fit these two branches to linear functions given by $S_{tot} = S_0^i - m_i p$, where the $S_{tot}$-axis intercepts and the slopes of these two linear functions are $S_0^i$ and $m_i$ respectively ($i$ = 1, 2). The transition pressure ($p_{trans}$) is given by

$$p_{trans} = \frac{S_0^1 - S_0^2}{m_1 - m_2} \quad (3)$$

The fitting parameters and the calculated p$_{trans}$ are given in **Table 1**.

**Table 1.** Transition pressure (p$_{trans}$) obtained from the linear fitting of total entropy vs pressure plot (**Error! Reference source not found.b**).

| T (K) | $S_0^1$ | $m_1$ | $S_0^2$ | $m_2$ | p$_{trans}$ |
|---|---|---|---|---|---|
| 250 | 38.0 | 0.31 | 30.5 | 0.07 | 31.1 |
| 300 | 51.4 | 0.38 | 36.2 | 0.06 | 47.3 |
| 350 | 54.2 | 0.32 | 37.2 | 0.05 | 63.2 |

Interestingly, the values of p$_{trans}$ obtained from here shows good agreement with those from diffusion (**Figure 2a** inset). This illustrates that the diffusion of water and the entropy of the system are strongly correlated. We further extrapolate the glassy branch of the entropy versus pressure plot towards lower pressures (dashed lines in **Figure 2b**) and subtract these values from the liquid state entropy [Eq. (4)]. This gives the excess entropy in the liquid with respect to the glass.

$$\Delta S = S_{liquid} - S_{glass} \quad (4)$$

At the transition pressure both $\Delta S$ and diffusion coefficient (D) become zero. In **Figure 2c** we plot D against $\Delta S$, which shows an exponential dependence [Eq. (5)].

$$D = Ae^{B\Delta S} \quad (5)$$

The exponential fitting parameters are given in **Table 2**.

**Table 2.** The exponential fitting parameters for the relationship between diffusion coefficient and excess entropy of liquid with respect to glass.

| T (K) | A ×10$^5$ (cm$^2$ s$^{-1}$) | B (J mol$^{-1}$ K$^{-1}$)$^{-1}$ |
|---|---|---|
| 250 | 0.006 | 0.69 |
| 300 | 0.06 | 0.24 |
| 320 | 0.03 | 0.27 |

From Eq. (5), we see that D = A when $\Delta S$ = 0. From **Table 2** we see that the values of A are close to zero. Hence, the mutual approach of D and $\Delta S$ towards 0 characterizes the vitrification process. It is fascinating to note that the exponential relationship observed in **Figure 2c** is similar to the well-known Rosenfeld diffusion-entropy scaling, although the definition of excess entropy is different.[48, 52]

## B. Density of States

The change in total entropy is mostly derived from the translational contribution (**Figure S2a**), which shows similar pressure dependence. However, rotational entropy remains almost constant throughout the pressure range (**Figure S2b**). This indicates that the rotational degrees of freedom are practically unperturbed with the increase in pressure. In a supercompressed state, the well-formed hydrogen bond network of water is compromised, as molecules are pushed closer to each other. Hence, subsequent to the pressure induced melting of ice Ih, the rotational degrees of freedom of water increases. This remains unaffected by further increase in pressure. However, the translational degrees of freedom suffer a substantial reduction, owing to the unavailability of free volume for movement. This is reflected in the decrease of translational entropy.

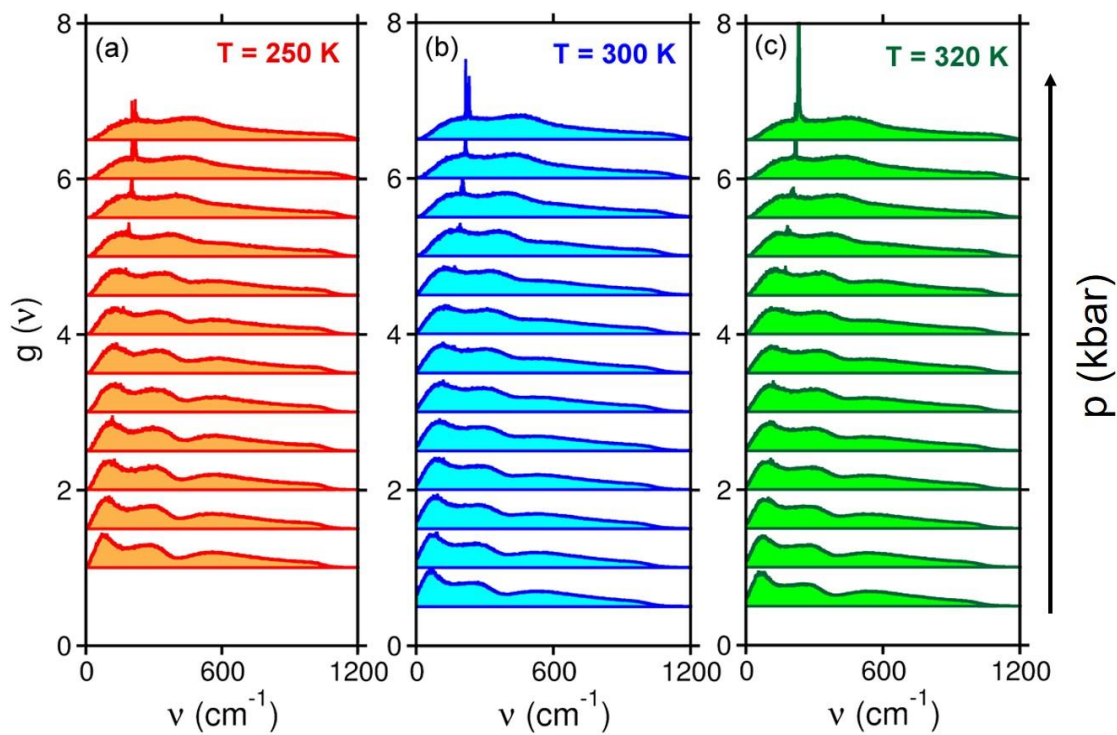

**Figure 3.** Change in the density of states [g(ν)] profile under the influence of pressure. With the increase in pressure, the frequency difference between the translational and rotational density of states decreases. Rotational degrees of freedom (~600 cm-1) does not show any significant pressure dependence, whereas, the translational counterpart (~60 cm-1 and ~200 cm-1) is noticeably modified. A sharp peak appears at ~230 cm-1 at very high pressure.

From **Figure 3**, we see that with the increase in pressure, the density of states profile changes significantly. The frequency difference between the translational (low frequency) and rotational (high frequency) degrees of freedom decreases, whereby, at very high pressures, they are no longer separated. We note that the broad peak at ~600 cm$^{-1}$ (rotational DOS) do not undergo any significant change, whereas the peaks at ~60 cm$^{-1}$ (O-O-O angle bend) and ~200 cm$^{-1}$ (O-O angle stretch) suffer substantial modification. At very high pressure, these two peaks are not separable. Furthermore, a sharp peak is found to appear at ~230 cm$^{-1}$, near the phase transition pressure. The origin of this peak is not clear and remains a subject of further research. This sharp feature is also found in the DOS of ice VII. Hence, the advent of this sharp peak marks the onset of the transition of the disordered state to a certain crystalline order, akin to ice VII.

## C. O-O-O Angle Distribution

Modification of the density of states of the O-O stretching and O-O-O bending modes occur because of the pressure induced reorientation of water molecules in configurational space. This is clearly captured in the O-O-O angle distributions for three neighbouring water molecules, as plotted in **Figure 4a**, **4b**, and **4c** for the 3 temperatures under consideration. The change in the shape of the O-O-O angle distribution shows a stark pressure dependence. In the molten disordered state, the distribution is broad and devoid of any particular feature. However, with the increase in pressure, the distribution starts to bifurcate to yield two distinct peaks near ~65° and ~115°. This feature is similar to the O-O-O angle distribution in ice VII (**Figure S3**). Therefore, this distribution holds signatures of the incipient transition of the glassy system to the crystalline ice VII. The emergence of the two-peak orientational structure is sharper at higher temperatures. As pressure is increased, the angular difference between these two peaks increases. We plot the change of this difference with pressure in at the four temperatures studied.

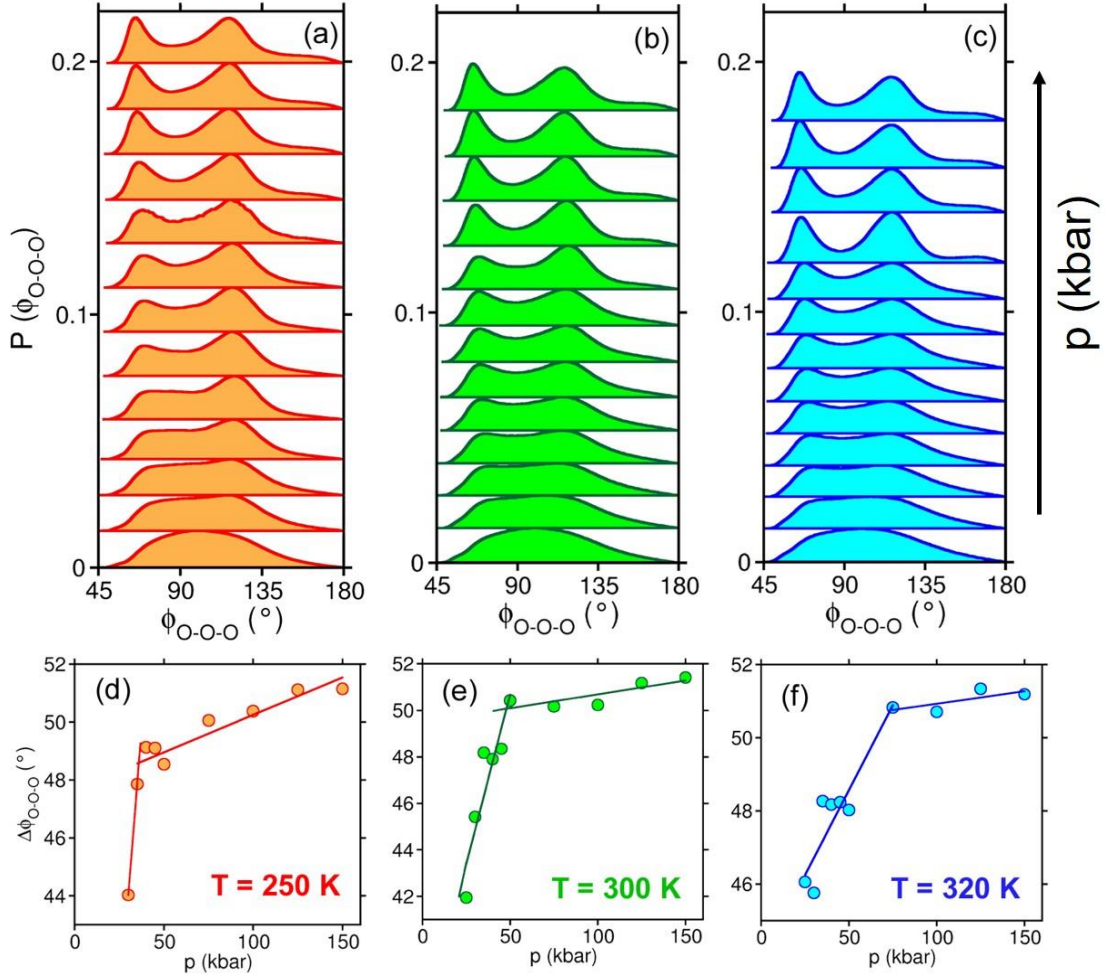

**Figure 4.** Distributions of O-O-O angles in the system at (a) 250 K, (a) 300 K, and (c) 320 K. Pressure increases from bottom to top. The highest pressure is 150 kbar while the lowest pressure is 10 kbar. The broad distribution in the molten state bifurcates into two distinct peaks (~65° and 115°) with the increase in pressure. This signifies a disorder to order transition. The peaks become sharper and the difference between them increases with increase in the compression of the system. This difference ($\Delta\phi_{O-O-O}$) is plotted at the 3 temperatures as a function of pressure in (d), (e), and (f) respectively. $\Delta\phi_{O-O-O}$ serves as an excellent order parameter that clearly captures the disorder to order transition at high pressure. The results are in good agreement with those obtained from diffusion coefficient and entropy.

In **Figure 4d**, **4e**, and **4f** the difference between the two peaks in O-O-O angle distributions ($\Delta\phi_{O-O-O}$) exhibits a sharp crossover near the transition pressure at all the three temperatures. We fit the plots to linear functions ($\Delta\phi_{O-O-O} = \Delta\phi_0^i + m_i p$) before and after the crossover. Here, $\Delta\phi_0^i$ and $m_i$ are the intercepts and slopes of two branches of the plot (I = 1, 2). We compute $p_{trans}$ similar to the way we use in the treatment of entropy [Eq. (6)]. The resultant data are presented in **Table 3**.

$$p_{trans} = \frac{\Delta\phi_0^1 - \Delta\phi_0^2}{m_2 - m_1} \quad (6)$$

**Table 3.** Transition pressure ($p_{trans}$) obtained from the linear fitting of $\Delta\phi_{O-O-O}$ vs pressure (p) plot (**Error! Reference source not found.d**, **5e**, and **5f**).

| T (K) | $\Delta\phi_0^1$ | $m_1$ | $\Delta\phi_0^2$ | $m_2$ | $p_{trans}$ |
|---|---|---|---|---|---|
| 250 | 21.04 | 0.77 | 47.66 | 0.026 | 35.8 |
| 300 | 36.12 | 0.29 | 49.49 | 0.012 | 48.1 |
| 350 | 43.92 | 0.09 | 50.25 | 0.007 | 76.1 |

Comparing $p_{trans}$ from **Table 1** and **Table 3**, we see that the estimates from entropy and from $\Delta\phi_{O-O-O}$ are very close to each other. Hence, $\Delta\phi_{O-O-O}$ serves as an excellent order parameter to detect the onset of the pressure induced recrystallization process. As we explain in the next section, this order parameter can successfully determine the pressure induced phase transition even at 80 K, where diffusion and entropy fail to do so.

### D. O-O Radial Distribution Function

In the highly compressed state, the molecules are pushed towards each other. This modifies the structural arrangements in the system significantly, which is distinctly captured in the oxygen-oxygen radial distribution function (RDF) [$g_{O-O}(r)$] of the systems. We plot the O-O RDF of the systems at the 3 temperatures and with increasing pressure in **Figure 5**.

The high density disordered state resulting from melting of ice Ih is structurally different from liquid water at ambient pressure. The presence of a shoulder peak at ~3.2 Å characterizes the high density of the system under compression. This peaks shifts towards the 1st sharp peak with increase in pressure. This shift is proportional to the change in density of the system (**Figure 1**). This peak ultimately merges with the 1st peak, making it broader.

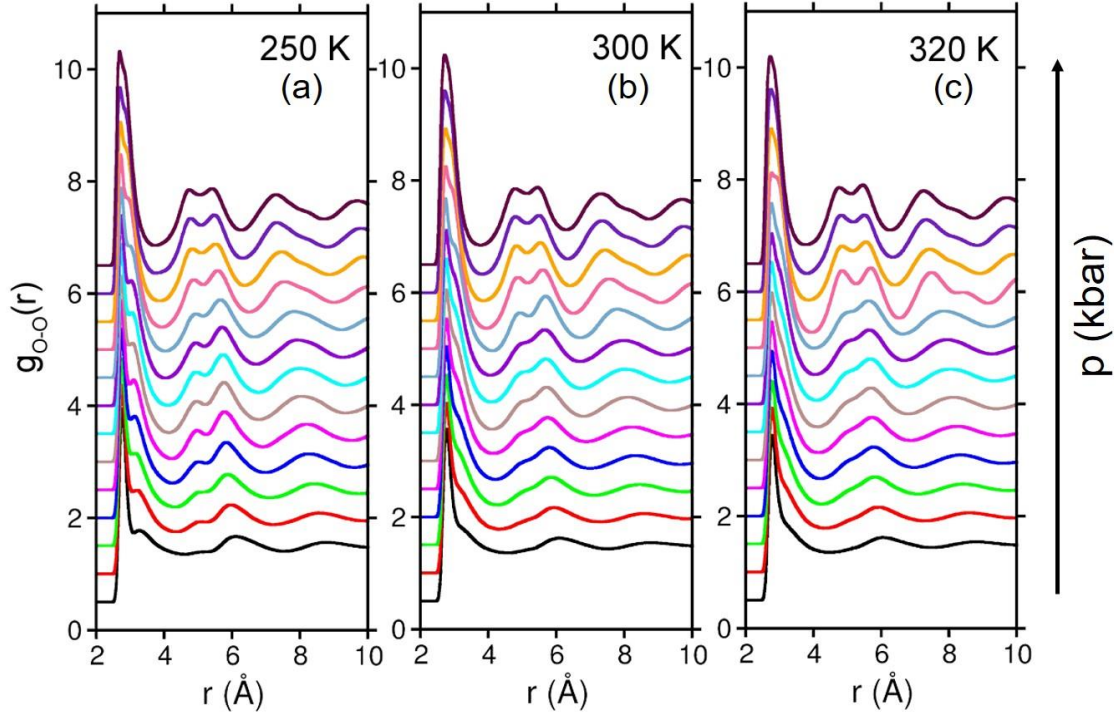

**Figure 5.** The structural changes denoted by Oxygen-Oxygen radial distribution functions [$g_{O-O}(r)$] at difference pressures at the 3 different temperatures (250 K, 300 K, and 320 K). With the increase in pressure the small peak adjacent to the first sharp peak shifts towards and ultimately coalesces with the latter. This suggests the decrease in compressibility of the system. With the increase in pressure, emergence of 2 peaks is observed at ~4.7 Å and ~5.4 Å. The structural development starts earlier at lower temperatures. These peaks are characteristics of ice VII. From 10 kbar to 150 kbar, the system traverses through a series of structural modifications, which are reminiscent of intermediate crystalline order.

Two new peaks start to emerge at ~4.7 Å and ~5.4 Å under the influence of pressure. These are signatures of crystalline order. Our results are in good agreement with the reports of Tse *et al.*[1] These two peaks are reminiscent of the structures of ice VII.

Analysing the nature of the RDF, we find that in the course of recrystallization, several highly entangled phases of ice, as reported in recent studies, appear.[22, 43] Therefore, our results agree with the observations of Yagasaki *et al.*[22] This process is akin to the Ostwald step rule, where a step-wise nucleation is observed during a freezing process. Just as Ostwald observed a long time ago, the formation of these intermediate ordered phases might be too quick and of too transient in nature to be observed in real experiments. The metastable phase, which is the most structurally similar to the parent crystal, freezes first.[59, 60] Interestingly, a recent experimental study indeed demonstrated that the thermodynamic region where VHDA is obtained, in fact, could house several crystalline polymorphs, sequential transitions among which lead to ice VIII at 100 K.[17, 45] This substantiates the view that the journey from ice Ih to ice VII involves several hidden exotic polymorphic phases of ice that are beginning to be deciphered and appreciated.

## III. Glass Transitions at 80 K

At 80 K, starting from ambient pressure and up to 150 kbar, we observe two glass transition phenomena. The 1$^{st}$ transition (ice Ih to high density amorphous ice (HDA)) at 13 kbar is associated with a sharp change in density. This is a well-documented observation.[1, 8, 28] However, during the second transition, the change in density is continuous. This is presented in **Figure 6a**.

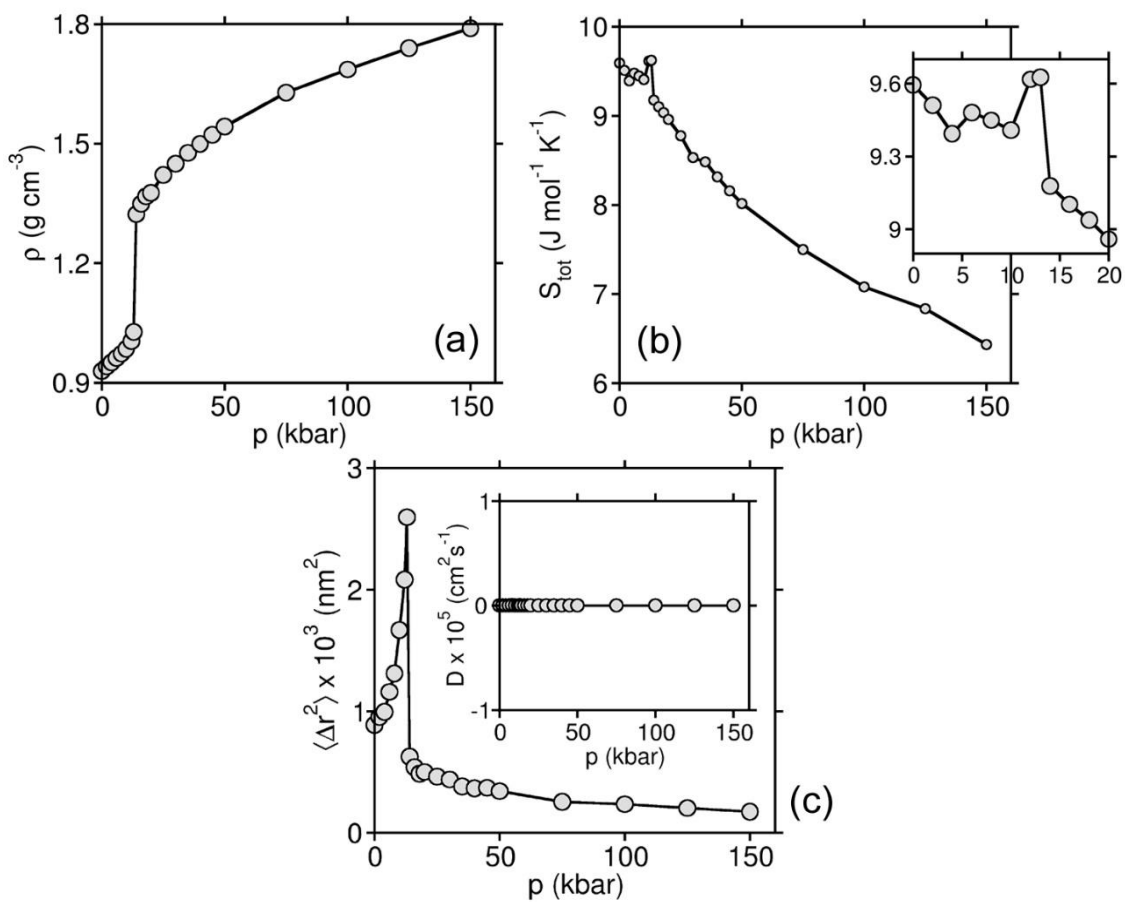

**Figure 6.** Pressure dependence of (a) density ($\rho$), (b) total entropy ($S_{tot}$), (c) mean square fluctuation (MSF) of molecular position of water and diffusion coefficient (D) (c inset) at 80 K. $S_{tot}$ experiences small, yet sharp decrease (zoomed view in the inset of b) at 13 kbar, where ice Ih crystal melts to high density amorphous ice (with ~30 % increase in density). With further increase in pressure, entropy decreases almost linearly, without any further crossover. MSF shows a sharp decrease at the transition pressure. It also does not hint at any subsequent phase transition. D remains 0 throughout the pressure range.

At the melting point (13.6 kbar) the system experiences a small yet sharp decrease in entropy (**Figure 6b**). Upon further compression, entropy decreases almost linearly, without any further crossover. The rate of decrease of entropy is much smaller than that observed at higher temperatures. The entropic contribution is derived completely from the harmonic oscillator partition function at 80 K [**Eq.(2)**] because there is no diffusive component. In the amorphous ice, the available free volume of individual water molecules, as advocated by the cell theory, decreases significantly with respect to the crystal. Consequently the vibrational degrees of freedom reduces and this results in a decrease of entropy.

In both the ice Ih crystal and the molten glassy phase the diffusion coefficient (D) of water molecules is zero. With further increase in pressure, the value of D remains zero, and does not reveal the formation of a new state at very high pressures (**Figure 6c** inset). Consequently, diffusion is unable to detect either of the two glass transitions at 80 K. Melting results in substantial decrease in the mean square fluctuation (MSF) $\left(\langle \Delta r^2 \rangle\right)$ of the positons of water molecules.

$$\langle \Delta r^2 \rangle = \frac{1}{N\tau} \sum_{i=1}^{N} \sum_{t=1}^{\tau} \left(r_i(t) - r_i^0\right)^2 \qquad (7)$$

Here, $N$ is the total number of water molecules in the system, $\tau$ is the total time, $r_i(t)$ is the instantaneous position of the $i^{th}$ molecule (O atom) at time $t$ and $r_i^0$ is a reference position (for example, the initial position of that molecule). We plot $\langle \Delta r^2 \rangle$ against pressure in **Figure 6c**. In the crystalline ice Ih state, the MSF increases with increase in compression. At the melting pressure, its experiences a sharp decrease when system enters the glassy phase. This justifies the entropic behaviour of the system. Thereafter MSF exhibits small decrease with pressure, without any further crossover. Hence, MSF distinctly determines the $1^{st}$ glass transition. However it fails to denote any subsequent transitions.

The pressure induced glass transition from ice Ih to HDA at 13 kbar is characterized by significant changes in the structural arrangement of water molecules. Increase in pressure beyond this transition point results in further structural and orientational reorganization at a molecular level. This is similar to the changes at higher temperatures (250 K, 300 K, and 320 K) as shown in **Figure 3** and **Figure 4**. The shoulder peak, adjacent to the $1^{st}$ sharp peak in O-O RDF shifts towards lower r values with the increase in pressure (**Figure 7a**). Unlike the high temperature scenario, these two peaks remain separable even at pressures as high as 150 kbar.

The two peaks near 4.7 Å and 5.4 Å are observed here as well. The low temperature region of the phase diagram, traversed in this study, ultimately leads to ice VIII (and not ice VII like the high temperature scenario). This observation is consistent with the early reports of Klein and coworkers.[1]

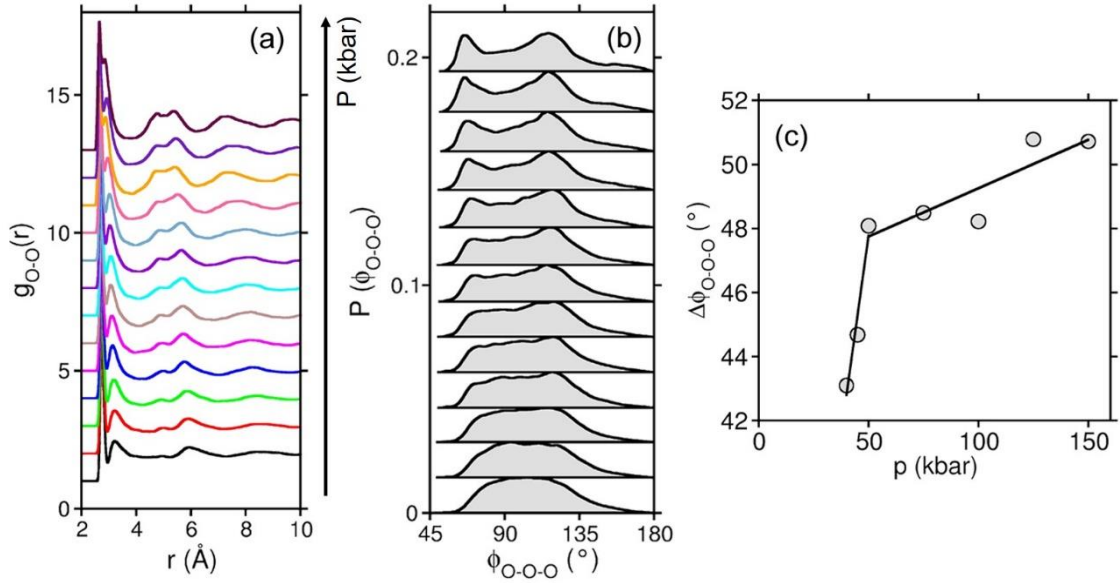

**Figure 7.** Pressure induced structural changes at 80 K. (a) O-O radial distribution function [$g_{O-O}(r)$], (b) O-O-O angle distribution, and (c) change in the difference between the two peaks in O-O-O angle distribution ($\Delta\phi_{O-O-O}$). Here, we observe a change in the slope of $\Delta\phi_{O-O-O}$ at p = 50 kbar, denoting a phase transition. The transition in this case is from one solid state to another. Both $g_{O-O}(r)$ and $\Delta\phi_{O-O-O}$ show signatures of crystalline order precursors at high pressure.

The pressure dependence of distribution of O-O-O angles also matches the observations at higher temperatures (**Figure 7b**). The bifurcation of the broad peak into two well-separated peaks denotes the presence of a disorder-order transition in the system, which is not observed in diffusion and entropy. Most notably, the difference between these 2 peaks ($\Delta\phi_{O-O-O}$) show a distinct crossover at 49.8 kbar (obtained in a similar fashion as before, using Eq. (6)), denoting a phase transition (**Figure 7c**). This is a solid-solid transition, which is different from the ones observed at higher temperatures, where a liquid state is transformed into a supercompressed glassy phase. Hence, the value of $p_{trans}$ does not follow the trend shown in **Figure 2** inset.

It is important to note that the glassy state obtained from the pressure induced vitrification is structurally different from the ones obtained at low temperatures and pressures. In that case, the distribution of O-O-O angles is broad and featureless, as in the liquid state.

Therefore we see that at 80 K, density and mean square fluctuation of molecular position can detect the 1st transition, but not the 2nd one. Diffusion coefficient shows no change through the pressure range from 0 to 150 kbar since at all pressures, the system is in a solid state. However, the 2nd glass transition pressure can be determined from the change in gradient of the difference between the positions of the 2 peaks in O-O-O angle distribution.

## IV. Conclusion

Let us briefly summarize the main results of this work. In this work we investigate the microscopic aspects of pressure induced glass transition in water. Ice Ih crystal is known to undergo melting or vitrification on compression, depending on the temeprature.[1, 2, 8, 27, 28] At relatively high temperatures, like 250K, the molten liquid water, on further increase in pressure, forms glass. On further compression, the glassy system shows structural modifications, with intriguing implications.[22, 43] However the microscopic state of the system, such as molecular arrangements, at such high pressures remains relatively less explored.

We study the effect of pressure on the disordered phase of water, which is obtained from the pressure induced melting of ice Ih at four different temperatures: 80 K, 250 K, 300 K, and 320 K. At 80 K the melting results in a crystal-to-glass phase transition giving a high density amorphous ice. The system experiences a sharp increase in density (~30 %). Here, the diffusion of water molecules remain zero. However, at higher temperatures (250 K, 300 K, and 320 K) the molten state is liquid, with non-zero diffusivity. With further increase in pressure, the diffusion becomes zero beyond a certain pressure, which is strongly temperature dependent. The static structure factor of the system in this state does not exhibit any crystalline order (**Figure S1**). Hence, at this point, the system experiences a liquid-to-glass transition. For the said 3 temperatures, the transition pressures ($p_{trans}$) are found to be 30 kbar, 50 kbar, and 65 kbar respectively. At 80 K, however, the diffusivity remains 0 throughout the pressure range from 0 kbar to 150 kbar and does not show any significant change during phase transition.

Diffusion exhibits a strong correlation with the entropy of the system. Pressure dependence of entropy exhibits a sharp crossover at the exact transition pressures, obtained from diffusion data. The majority of the total entropy is derived from the translational contribution, which itself shows similar pressure dependence. However, the rotational counterpart remains unperturbed under the influence of pressure, with negligible change (**Figure S2**). Diffusion coefficient exhibits an exponential dependence on the excess entropy of the system (as compared to the glassy state), akin to the Rosenfeld diffusion-entropy scaling.

At 80 K, the initial melting transition shows a small but sharp decrease in entropy because of the decrease in the available free volume of the system. The changes in entropy are primarily derived from the changes in the density of states. We find that the rotational modes (~600 cm$^{-1}$) do not show any significant change with increase in pressure. However, the translational modes (O-O-O bending at ~60 cm$^{-1}$ and O-O stretching at ~200 cm$^{-1}$) shows substantial shifts. At very high pressures, the difference between these modes almost vanishes.

Perturbation of the said translational modes results in reorganization of molecular orientation, which is clearly captures in the distribution of O-O-O angles between 3 neighbouring water molecules. At lower pressures (below 30 kbar), this distribution is broad and structureless. However, with the increase in pressure, the distribution starts to bifurcate, giving 2 well separated peaks at ~65° and 115°. This is only possible, when a certain crystalline order appears in the system. The observed 2-peak structure of O-O-O angle distribution is akin to the same in ice VII. Hence, we get indications of incipient crystalline order under the influence of pressure. Most interestingly, when plotted against pressure, the difference between the positions of the two peaks ($\Delta\phi_{O-O-O}$) shows a change in slope at the transition pressures obtained from diffusion and entropy. Hence, $\Delta\phi_{O-O-O}$ serves as a good order parameter to detect the pressure induced glass transition.

Modification in structural order is also observed in O-O radial distribution function of the system with increase in pressure. Most notably, the small peak, which appears because of the high density of the system, shifts towards the 1$^{st}$ sharp peak with increase in pressure. This suggests an increase in the density of the system. At 80 K, these two peaks remain separated, whereas at higher temperatures, they merge with each other. Two new peaks originate at ~4.7 Å and ~5.4 Å, which again shows correspondence with the O-O RDF of ice VII. These are precursors of crystalline order in the system. The characteristics of the O-O RDF agrees with the reports of Yagasaki *et al.*[22]

Energetics of the system might play an important role in the course of the observed phase transition. Structure and thermodynamics of a molecular system are determined by the requirement of free energy minimum. In the present problem, this is complicated by the interplay between the short-range Lennard-Jones and the long-range electrostatic interactions between the water molecules. It is well known that Argon (described by LJ interaction potential) easily crystallizes to FCC (Face-Centred-Cubic) lattice.[61, 62] With the intervention of electrostatic forces (Coulomb interaction), requirement of hydrogen bonding start dictating the free energy. However, at a density much larger than that of ice Ih, our conventional description of water in terms of hydrogen bonding becomes inadequate. Nevertheless the electrostatic interactions continue to play an important role. Under extreme compression when molecules are pushed closed to each other, the effect of LJ potential becomes important, and close packing corresponding to certain crystal structures becomes admissible from free energy minimization. Nonetheless, it is nontrivial to obtain a perfect crystal because of the necessity to navigate strong Coulomb interactions. Hence, we believe that this competition lies at the heart of the problems addressed in this work and could possibly explain the origin of the large number of ice polymorphs observed at high pressure. A detailed probe into the energetics of the system is required to clearly grasp the role of this interplay in the observed phase transitions.

The phenomenon investigated in this work holds an interesting analogy with the random spin glass science. The analogy helps explain the fast nucleation of ice VII from VHDA. In the language of random spin glass, the phases that are of intermediate order between the metastable parent phase and the final stable daughter phase discussed here could wet the interface between HDA and ice VII, thus lowering the surface tension between the two phases. The wetting scenario has been used in the theory of glass transition.[63-65] The analogy between spin glass and Ostwald step rule is indeed tantalizing and deserve further study.[29, 30]

## V. Simulation Details

The initial structure of the ice Ih is generated using the GenIce package.[66] The crystal contains 1024 water molecules. We use the TIP4P/Ice model[67] of water and perform molecular dynamics simulations in GROMACS-5.1.4 MD engine.[68] The initial structures are subjected to thorough energy minimization using a succession of steepest decent and conjugate gradient algorithms.[69] Firstly, the ice Ih crystal is molten separately at four different temperatures (T = 80 K, 250 K, 300 K and 32 K) by compression. Subsequently, we equilibrate the systems for 100 ns each, separately at the given temperatures and pressures (NPT conditions). At 250 K, 300 K and 320 K, the studied pressures are 10, 15, 20, 25, 30, 35, 40, 45, 50, 75, 100, 125, and 150 kbar. At 80 K, we simulate the system at pressures of 0, 2, 4, 6, 8, 12, 13, 14, 16, and 18 kbar, along with the pressures mentioned before. Ice VII is simulated separately at T = 250 K and P = 150 kbar for comparison with the other system.

The temperature and pressure are controlled using the Nose-Hoover thermostat ($\tau_t$=0.21 ps) [70, 71] and Parrinello-Rahman barostat ($\tau_p$=0.1 ps) [72] respectively. We perform the production runs on the equilibrated structures for 1 ns at the given temperature and pressure, with 100 fs data dumping frequency. For calculation of entropy using the 2PT method,[55-57] we run separate 20 ps simulations with data dumped at 4 fs intervals. We run the simulations using the leap-frog integrator algorithm [73, 74] with dt = 1 fs. The analyses are performed on these trajectories. The cut-off radius for neighbour searching and non-bonded interactions are taken to be 10 Å.

For the calculation of electrostatic interactions we use the Particle Mesh Ewald (PME) algorithm[75] with an FFT grid spacing of 1.6 Å.

## Acknowledgement

The simulations were performed in the Cray Supercomputer 'SahasraT' stationed at the SERC (Supercomputer Education and Research Centre) Department of IISc. BB thanks DST, India, for partial funding of this work, and SERB, India for providing National Science Chair (NSC) Professorship. SM thanks DST, India for INSPIRE fellowship and IISc for Institute Research Associateship.

# Supplementary Figures

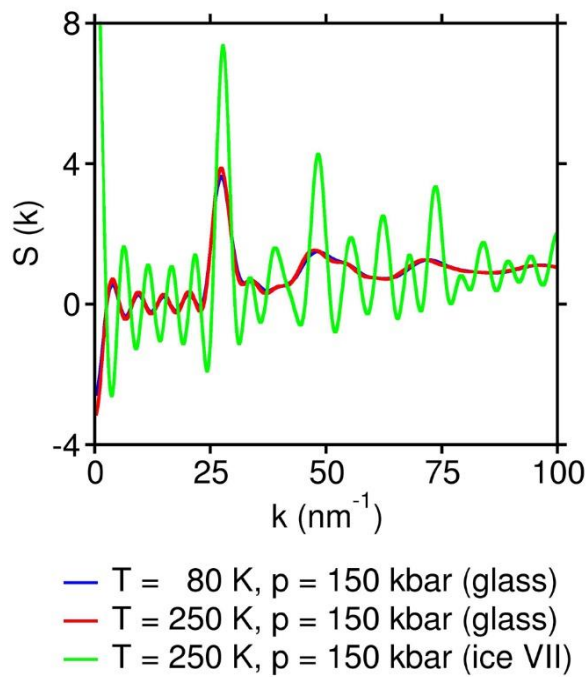

**Figure S1.** Static structure factor [S(k)] of the phase of water (at p = 150 kbar) formed by compression of the disordered state generated from the pressure induced melting of ice Ih at 80 K (blue) and 250 K (red). These are compared with the same for ice VII at p = 150 kbar and T = 250 K (green). The S(k) of the system at the two temperatures are same. It does not display any characteristics of a crystal. The diffusion of water under these conditions is zero. Combining these dynamical and structural features, these compressed phase of water behave like a glass.

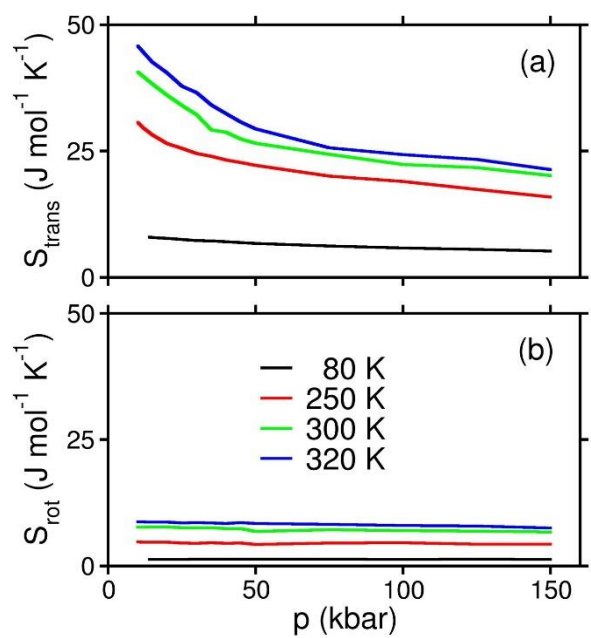

**Figure S2.** (a) Translational entropy ($S_{trans}$), and (b) Rotational entropy ($S_{rot}$) are plotted as functions of pressure. The major fraction of total entropy ($S_{tot}$) is derived from $S_{trans}$. $S_{trans}$ shows a crossover at the transition pressures at 250 K, 300 K and 320 K. At 80 K, however, there is no change of slope. $S_{rot}$, on the other hand, remains practically constant throughout the pressure range at all the temperatures.

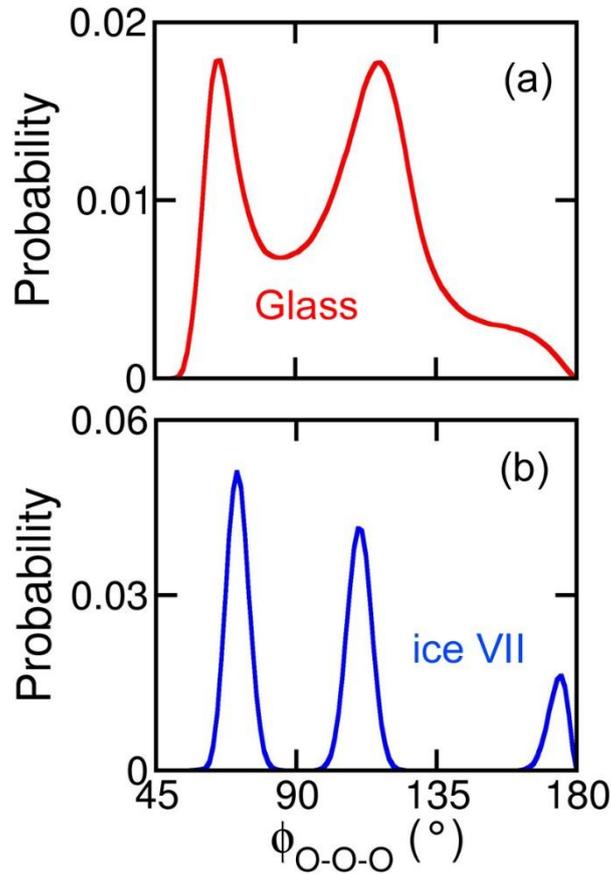

**Figure S3.** O-O-O angle distribution in (a) the compressed glassy phase and (b) ice VII at p = 150 kbar and T = 250 K. The distribution in (a) shows stark similarity with that in (b). The appearance of the two peaks at ~65° and ~110° in the glassy phase at high pressure suggests the subsequent formation of ice VII from this phase.